\documentclass[%
 aip,
 amsmath,amssymb,
 reprint,%
]{revtex4-1}
\usepackage[dvipsnames]{xcolor}

\usepackage{graphicx}
\usepackage{dcolumn}
\usepackage{bm}

\usepackage[utf8]{inputenc}
\usepackage[T1]{fontenc}
\usepackage{mathptmx}
\usepackage{etoolbox}

\makeatletter
\def\@email#1#2{%
 \endgroup
 \patchcmd{\titleblock@produce}
  {\frontmatter@RRAPformat}
  {\frontmatter@RRAPformat{\produce@RRAP{*#1\href{mailto:#2}{#2}}}\frontmatter@RRAPformat}
  {}{}
}%
\makeatother
\begin{document}

\preprint{AIP/123-QED}

\title[A Compact Low-level RF Control System for ACCEL]{A Compact Low-level RF Control System for Advanced Concept Compact Electron Linear Accelerator}
\author{C. Liu}

 \email{chaoliu@slac.stanford.edu}
 
\author{L. Ruckman}%
\author{R. Herbst}%
\author{B. Hong}%
\author{Z. Li}%
\author{K. Kim}%
 \affiliation{ 
SLAC National Accelerator Laboratory, Menlo Park, California, USA.
}%

\author{D. Amirari}%
\author{R. Agustsson}%
 \affiliation{ 
RadiaBeam Technologies LLC, Santa Monica, USA.
}%

\author{J. Einstein-Curtis}%
\author{M. Kilpatrick}%
\author{J. Edelen}%
 \affiliation{ 
RadiaSoft LLC, Boulder, USA.
}%

\author{E. Nanni}%
\author{S. Tantawi}%
\author{M. Kemp}%

 \affiliation{ 
SLAC National Accelerator Laboratory, Menlo Park, California, USA.
}%

\date{\today, The following article has been accepted by Review of Scientific Instruments by the AIP.}

\begin{abstract}
A compact low-level RF (LLRF) control system based on RF system-on-chip (RFSoC) technology has been designed for the Advanced Concept Compact Electron Linear-accelerator (ACCEL) program, which has challenging requirements in both RF performance and size, weight and power consumption (SWaP). The compact LLRF solution employs the direct RF sampling technique of RFSoC, which samples the RF signals directly without any analogue up and down conversion. Compared with the conventional heterodyne based architecture used for LLRF system of linear accelerator (LINAC), the elimination of analogue mixers can significantly reduce the size and weight of the system, especially with LINAC requires a larger number of RF channels. Based on the requirements of ACCEL, a prototype LLRF platform has been developed and the control schemes have been proposed. The prototype LLRF system demonstrated magnitude and phase fluctuation levels below 1\% and  1\textdegree \,on the flat top of a 2 \(\mu\)s RF pulse. The LLRF control schemes proposed for ACCEL are implemented with a prototype hardware platform. This paper will introduce the new compact LLRF solution and summarize a selection of experimental test results of the prototype itself and with the accelerating structure cavities designed for ACCEL.
\end{abstract}

\maketitle

\section{Introduction}

The Advanced Concept Compact Electron Linear-accelerator (ACCEL) program aims to develop a high-power compact, rugged linear accelerator that can be easily transported and deployed in the field. The most critical and challenging requirement of the ACCEL is the size, weight and power consumption (SWaP) of the linear accelerator (LINAC) \cite{accel}. All the RF components of ACCEL, including the RF source, amplification, acceleration structure, and low-level RF (LLRF), have been designed to achieve high RF power with stringent SWaP requirements. In this paper, the prototype compact LLRF system for ACCEL, including the hardware platform and control schemes, will be introduced and discussed. 

The ACCEL system has been designed to operate in the C-band around 5.712 GHz. The accelerating structure has 26 pairs of cavities, and with each pair shares a single RF input and two RF monitors. With the conventional LLRF system with heterodyne based architecture, at least 78 analogue up and down mixing circuits with discrete data converters would be required. The size and weight of the LLRF hardware will be substantial after including the field programmable gate array (FPGA) and processing system (PS) for control and interfacing. The compact LLRF system for ACCEL has been designed based on the RF system-on-chip (RFSoC) device family from AMD Xilinx. The RFSoC device integrates RF data converters, FPGA and PS, which are the essential components of an LLRF system. The integrated data converters in RFSoC devices offer higher-order Nyquist zone sampling, enabling C-band RF signals to be sampled or generated directly without analog mixing. The RFSoC device selected for compact LLRF, AMD Xilinx Zynq UltraScale+ RFSoC Gen 3 ZU49DR, has 16 RF analogue to digital converters (ADCs) with maximum RF input frequency of 6 GHz and a maximum sampling rate of 2.5 GHz and 16 RF digital to analogue data converters (DACs) with a maximum sampling rate of 9.85 GHz \cite{rfsoc}. The LLRF system for the complete ACCEL structure can be realized with 4 RFSoC devices with a minimum of other electronics. The high integration level of the RFSoCs offers substantial benefits in the overall SWaP of the ACCEL.
\begin{figure*}
\includegraphics[width=7 in]{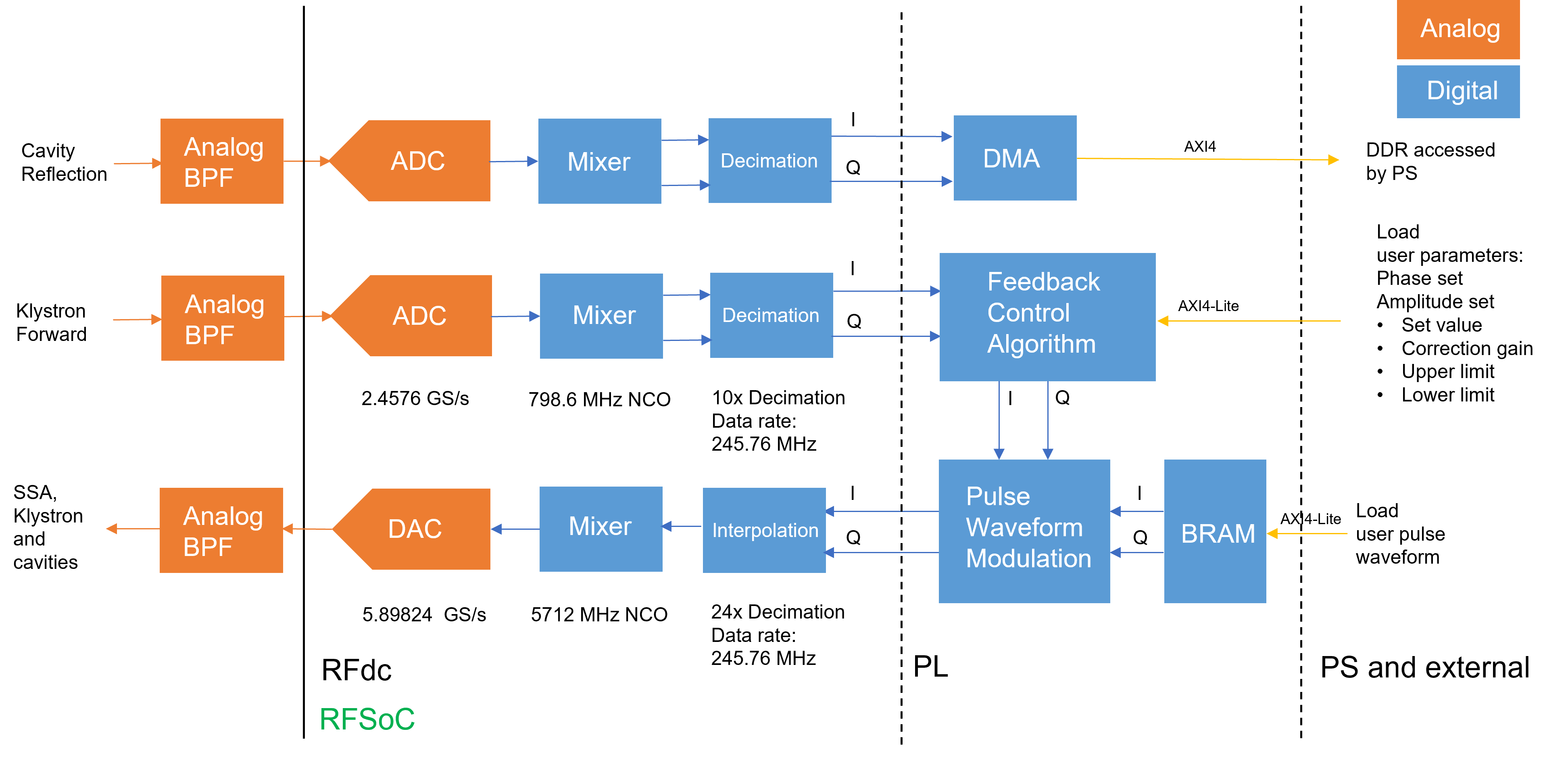}
\caption{\label{fig:fig-1} The block diagram of the compact LLRF platform designed for ACCEL. A band pass filter (BPF) is used to reject the noise outside of the frequency range of interest. Data is transferred to the processing platform using direct memory access (DMA), which provides the high bandwidth access between different interfaces. Custom waveforms are stored in on-chip block random access memory (BRAM) to allow for custom pulse shapes.}
\end{figure*}

RFSoC technology has been widely used in the research and development of control and readout platforms for a variety of physics experiments at the SLAC National Accelerator Laboratory and other institutes with which we collaborate \cite{liu2021characterizing, liu2022development,  liu2023evaluating, henderson2022advanced, liu2023higher,liu2024development}. The LLRF for the cool cooper collider (C\(^3\)), which also operates in the C-band, has been prototyped based on an RFSoC device in the same family as part of the road map for C\(^3\) R\&D \cite{Bai:2021rdg,Vernieri:2022fae,nanni2023status,andorf2025esppu}. The initial RF performance evaluation of the LLRF for C\(^3\) with a solid state amplifier (SSA) demonstrated that the magnitude and phase fluctuation levels within 1 s are 0.34 \% and 0.37\textdegree \,respectively \cite{liu2024direct}, which is considerably better than the requirements of 1\% and 1\textdegree \ for ACCEL. In \cite{liu2024generationllrfcontrolplatform}, the next generation LLRF (NG-LLRF) platform based on RFSoC developed at SLAC has been introduced. The NG-LLRF platform demonstrated that it can drive and measure a high-power test setup with high precision in both the C-band \cite{liu2024direct,liu2025high} and the S-band \cite{liu2025next}. The compact LLRF system for ACCEL has been designed and implemented based on configuration optimization and performance characterization exercises performed for RFSoCs with the other LLRF systems of LINACs and other physics experiments. In Section \ref{sec:platform}, the architecture of the prototype compact LLRF system, including hardware, firmware, and software, will be introduced.

An LLRF system typically employs a pulse-to-pulse feedback loop to stabilize the RF field in an accelerating structure \cite{geng2017rf}. The proposed pulse-to-pulse magnitude and phase control scheme will be described in Section \ref{sec:rf_ap}. Due to the special operating conditions of ACCEL, more control schemes are required. One of the design requirements of the ACCEL is to operate in a wide temperature range without temperature stabilization for the accelerating structure. The resonant frequency of the accelerating structure can drift with temperature, which can affect the injection of RF power into the structure. We have designed an RF frequency tuning scheme to track the resonance of the structure cavities and change the RF frequency accordingly. The RF frequency tuning scheme will be described in Section \ref{sec:rf_freq}. The ACCEL system is designed to operate with an electron bunch train within an RF pulse, requiring the phase and magnitude to be stable within each pulse. The scheme proposed to control the magnitude and phase within the pulse will be described in Section \ref{sec:rf_flatness}.

The compact LLRF prototype has been tested in some critical stages of t development. The first test with the prototype has been performed is the direct loopback test with platform and the fluctuation levels of the loopback setup define the highest stability level of entire control system. The test results of the loopack test setup will be summarized in Section \ref{loopback}. The RF frequency tuning scheme has been implemented and tested with the ACCEL structure cavities. The phase and magnitude of the cavity reflection signals before and after the tuning scheme applied will be demonstrated and discussed in Section \ref{tuning}. A prototype of the compact LLRF system has also been tested at the high-power C-band test facility with a water load. A selection of test results will be summarized in Section \ref{highpower}.

\section{\label{sec:platform}Compact LLRF Control Hardware Platform}

The compact LLRF prototype for ACCEL has been developed based on the ZCU216 evaluation board, which carries an RFSoC Gen 3 ZU49DR device. A block diagram for the platform is shown in Figure \ref{fig:fig-1}. The solid line in Figure \ref{fig:fig-1} defines the boundary between the RFSoC and the external components and signals. Within the RFSoC, the system implementation has been divided into three parts separate by the dashed lines: the RF data converter (RFdc), the programmable logic (PL) and the processing system (PS). The mixers and decimation or interpolation blocks before or after the data converter are integrated as a part of RFdc and each of the chains is named as a `datapath'. The datapaths for the data converters can be configured independently of each other.

For each pair of the ACCEL cavities, there are two RF signals to be measured and used for control and monitoring purposes, which are the klystron forward signal and the cavity reflection signal. As Figure \ref{fig:fig-1} shows, RF signals are filtered with a band-pass filter and  then sampled directly by the integrated ADCs in RFSoC at 2.4576 giga samples per second (GSPS). The digitized signal is then down converted to 798.6 MHz, which is the image of 5.712 GHz in the first Nyquist zone. The base-band signal in in-phase (I) and quadrature (Q) format is decimated by a factor of 10. Since the signal is quadrature sampled after the digital down-mixing, the bandwidth of the RF measurement chain in this case is 245.76 MHz, which is the same as the data rate.  The bandwidth can be adjusted for future requirements using a different decimation factor. The device uses digital finite impulse response (FIR) filters to reject high frequency noise as a part of the decimation process.

For the ACCEL structure design, there are no probes in the cavity structure. The cavity reflection is the only signal that carries information about the field inside the structure. In this version of the prototype, the IQ components of the cavity reflection signal in streaming format have been converted to memory-mapped format and written to the DDR, which is outside of the RFSoC. The data can then be accessed by the PS and sent to a server or processed locally on the PS. For the experiments in this paper, the data was sent to a server via Ethernet for further processing.  

The down-converted klystron forward signal in IQ format is streamed to the feedback control algorithm firmware block. The user input parameters from the software layer are loaded to the firmware block via registers. Based on the IQ components and user input parameters, the control block calculates the updated IQ pair that compensates for the pulse-to-pulse fluctuation in phase and magnitude levels. The user can load a custom waveform via software; otherwise the default square wave will be used. The waveform is modulated with the updated IQ in the modulation block, which generates the updated base-band pulse. The DAC datapath interpolates the base-band pulse first and then up mixed with the digital values from the numerical controlled oscillator (NCO). The digital up conversion frequency can also be controlled via software. The DAC decodes and generates the updated RF pulse with a sample rate  of 5.89824 GHz and drives the entire structure.

For the ZCU216 platform we selected, the master clock for the RFSoC is generated by an external phase locked loop (PLL) device from a reference signal. In this case, we have only one RFSoC, so an on-board 10 MHz oscillator is used as a reference. However, when there are multiple RFSoCs in the system, all devices need clocks that are frequency locked and phase aligned.  For each of the RFSoCs, the master clock is delivered to one of the data converter tiles and propagated to other tiles using an internal network. The clocks for sampling, NCO and filtering blocks of each tile are generated by internal PLLs from the master clock propagated to the tile. Therefore, all integrated blocks of the RFSoC devices are synchronized with the master clock. The propagation delays between the data converters and the programming logic can be handled by the Multiple Tile Synchronization (MTS) technique of the RFSoC family. With MTS, the latency between tiles can be deterministically aligned between reset and reset. Reset of NCOs from different tiles or different devices should be triggered by a common system reference to achieve repeatable operation.


\section{\label{sec:level1}LLRF Control Schemes}

The ACCEL LLRF control system has been divided into three main parts based on the requirements. The control schemes proposed are designed to be performed in order: RF frequency control, followed by pulse shape control, and then the pulse-to-pulse phase and magnitude control feedback loop. The control schemes proposed are demonstrated in the form of flow charts in the following three subsections. In the flow charts, the orange blocks are planned to be implemented in software and the blue blocks to be implemented in firmware. The interface between the software and the firmware will be defined in Section \ref{sec:interface}.

\subsection{\label{sec:rf_freq}RF frequency tuning}

The target operating temperature range of ACCEL is from -40 to 85 \textdegree C. The resonant frequency of the accelerating changes by around 0.1 MHz per \textdegree C. Due to the limits in size and weight, there is no temperature stabilization mechanism for the accelerating structure of ACCEL and maintaining beam and beam energy is the priority of the program. Therefore, the RF frequency needs to be tuned according to the resonance frequency of the structure while operating. 

\begin{figure}
\includegraphics[width=3.4in]{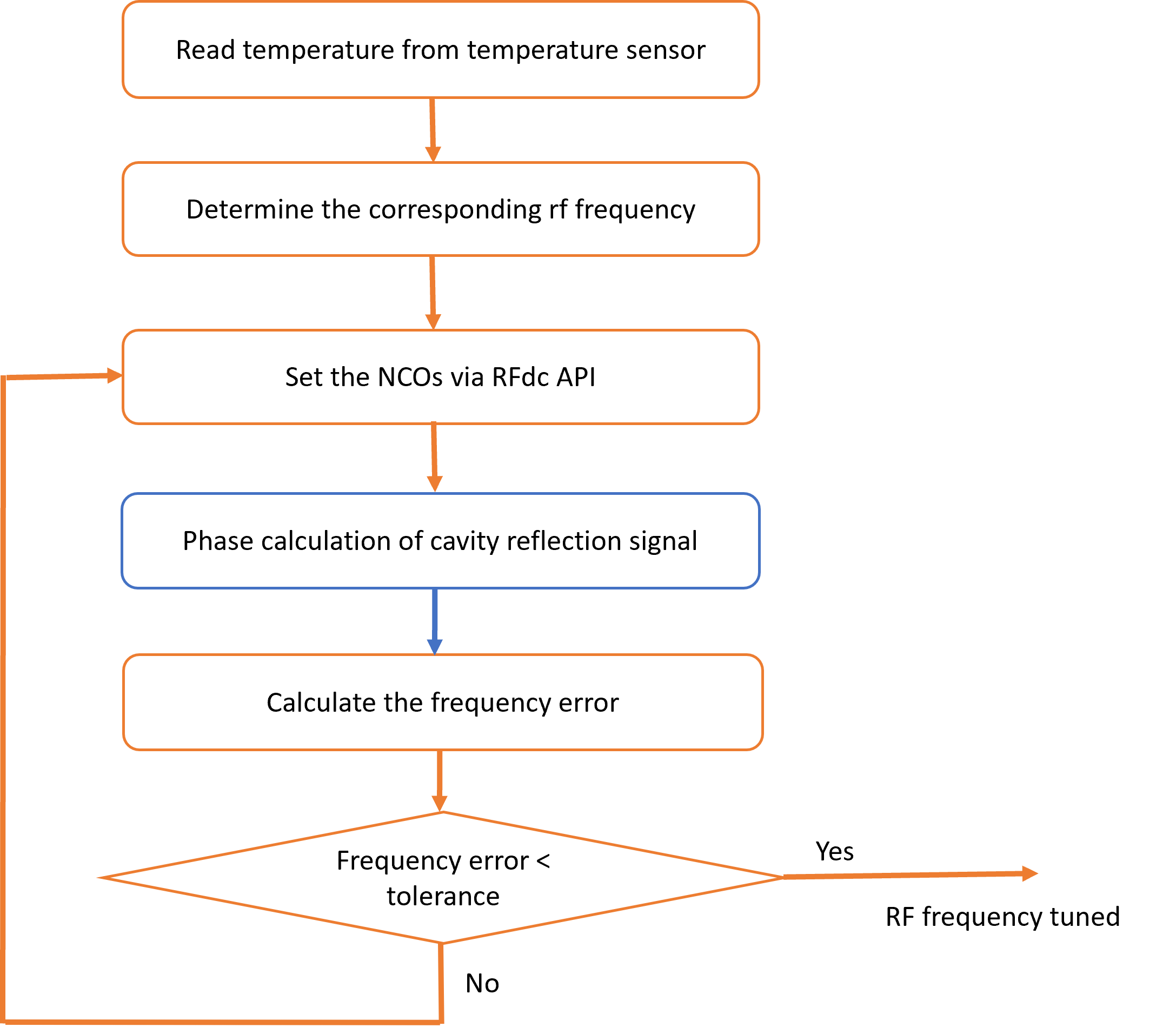} 
\caption{\label{fig:fig-2} The flow chart of the RF frequency tuning routine proposed for ACCEL LLRF system.}
\end{figure}

\begin{figure}
\includegraphics[width=3.4in]{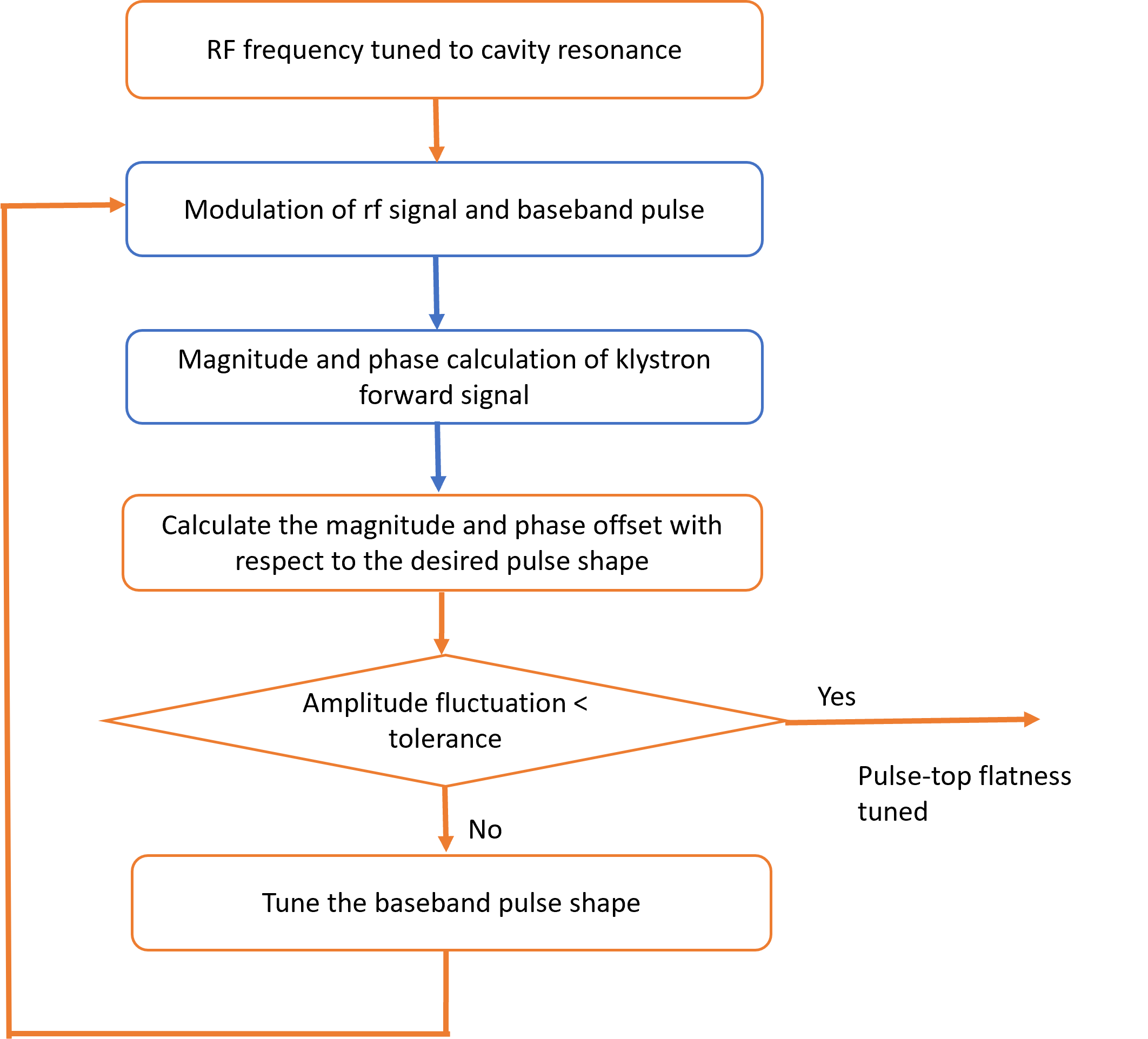} 
\caption{\label{fig:fig-3} The flow chart of the RF shape control routine proposed for the ACCEL LLRF system.}
\end{figure}

Figure \ref{fig:fig-2} shows the flow chart of the RF frequency control flow. When the system is first switched on, the RF frequency tuning flow will be executed. The RF frequency will be coarsely set based on the reading of the temperature sensor and the lookup table of the temperature versus RF frequency embedded in the software. With the coarse RF frequency tuning, a fraction of RF power should be able to be injected into the cavity, and the energy stored in the structure will radiate at the resonant frequency of the structure after the RF pulse is off. The radiation can be captured by the cavity reflection coupler, and the IQ samples of the cavity reflection signal from the ADC datapath are sent to software. The frequency error can then be calculated based on the phase of the cavity reflection signal after the RF pulse is off, and the calculation theory and experimental results will be elaborated in Section \ref{tuning}. The updated RF frequency calculated by the software is loaded into the datapaths of the data converters and changes the frequency of the sequence generated by the NCOs, which will ultimately update the RF frequency for both the ADCs and DACs. The frequency correction process can be executed in a controlled step size until the frequency error is below a desired value defined by the user. The RF frequency tuning will be performed each time the structure is switched on. But it will also be executed when the ambient temperature changes and introduces a detectable structure resonance frequency change above a user-defined threshold. Based on the gradient of temperature change during operation, the frequency tuning loop can be executed at an adaptive rate.

\subsection{\label{sec:rf_flatness}RF pulse shape control}

Figure \ref{fig:fig-3} shows the flow chart of the pulse shape control scheme. After the RF frequency tuning scheme is applied to the system, the pulse shape control flow will be executed. As there is no cavity probe that measures the RF field in the structure directly and the RF field follows the klystron by design, the cavity forward and reflected signals could be used to determine the field strength in the cavity and control the needed shape of the RF pulse. The magnitude and phase values of the klystron forward signal are sent to the software to calculate the updated pulse shape that can compensate for any fluctuation within the pulse. In software, the updated pulse shape is calculated based on the variation between the desired and measured pulse shape to reduce the offset between them sample by sample. Then the updated base-band pulse shape in IQ format is loaded to the firmware and up-converted to drive the system for the following pulse. The procedure repeats until the standard deviations of magnitude and phase on the flat top are below the desired levels. The pulse shape should be consistent after the compensation procedure is performed, so the pulse shape control may not require being executed continuously after the first run. In \cite{liu2025high,liu2025highipac}, we summarize the test results of different RF pulse modulation schemes with an LLRF system based on RFSoC integrated with high-power test stands in the C-band. The system demonstrates high precision in the generation and measurement of RF pulses with amplitude and phase modulation schemes. However, operating the accelerator with an arbitrary waveform is the first step of pulse shaping, and more development and test efforts are required for the control routine and algorithm to deliver the field with optimum stability to the accelerating structure. Since the pulse shape may change with the drifts of the RF components, the stability of the control scheme should be evaluated in the next phase of development.

\subsection{\label{sec:rf_ap}Pulse-to-pulse Magnitude and Phase Control}

\begin{figure}
\includegraphics[width=3.4in]{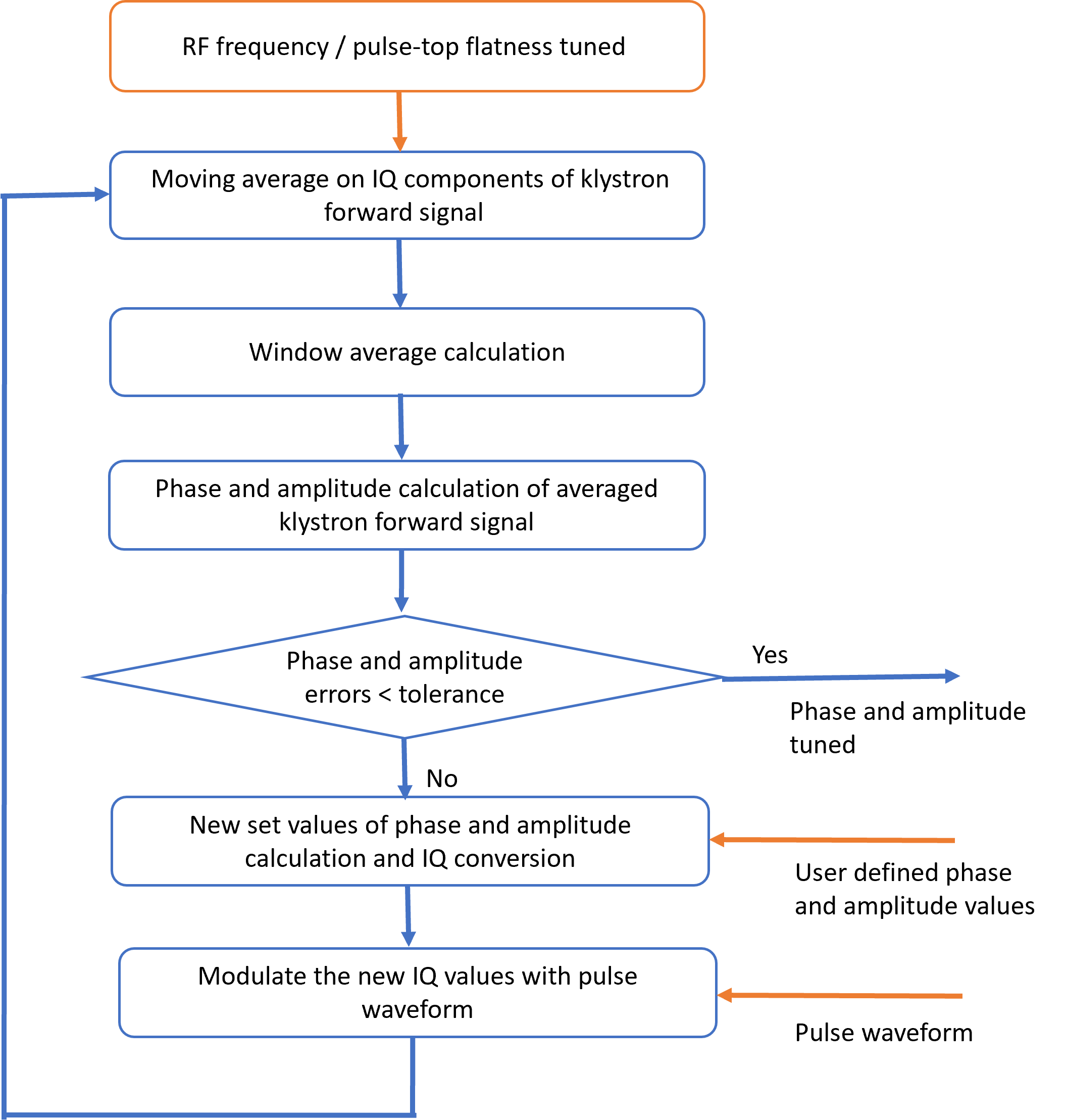} 
\caption{\label{fig:fig-4} The flow chart of the pulse-to-pulse magnitude and phase feedback control loop proposed for the ACCEL LLRF system.}
\end{figure}

Measurement and stabilization of the magnitude and phase of the field in the accelerating structure are the primary objectives for most of the LLRF control systems of LINACs. Figure \ref{fig:fig-4} shows the flow chart of phase and magnitude control that we have proposed for ACCEL. After the RF frequency and pulse-top flatness are tuned to desired levels, the phase and magnitude feedback control loops will start operating. This control loop is also be implemented by using the klystron forward signal. In firmware, the base-band klystron forward signal from the ADC datapath will be processed with moving average with a certain number of samples to improve the signal stability. The IQ components are converted to phase and magnitude values, and the average values will be computed. The current average magnitude and phase values will be compared to the user defined values and then calculating for a new set of magnitude and phase values to approach the user defined values in the defined rates. The new set of magnitude and phase values is converted back to IQ format and modulated with the base-band pulse waveform. The up converted RF signal samples are finally clocked into the DAC to generate the updated RF pulse, which will be used as the drive of the accelerating structure for the following pulse. The feedback loop runs continuously during operation to perform real-time pulse to pulse magnitude and phase correction. If the difference between the measurement and the desired value is below a certain threshold, the values will not be changed for the following pulse. This is done to minimize the instability introduced by the LLRF if the desired signal levels are already meeting the operating requirements.

\subsection{\label{sec:interface}Interface between Firmware and Software}

The data converter and firmware implemented in RFSoC is managed by a suite of software based on SLAC’s Rogue architecture \cite{rogue}, which provides the hardware abstraction layer and interface to higher level software. The low-level interface between the firmware and software is facilitated by a custom kernel driver that provides an interface to streaming data through a firmware DMA engine as well as providing an application programming interface (API) for reading and writing registers contained in the FPGA.  

The interface for the communication between the hardware abstraction layer and the higher-level software is implemented with the Experimental Physics and Industrial Control System (EPICS) version 7 process variable access (pvAccess) variables \cite{epics} while also having hooks for integration with other command protocols.Base-band samples will be presented as Numpy arrays of the appropriate data type and scalar variables are presented as their native type. The state control, state monitoring, and other interfaces between higher level control software and the hardware abstraction layer will also occur over EPICS V7. A client interface os provided to access a debug interface directly to the firmware and associated support hardware. Additionally, the hardware abstraction layer facilitates streaming data for debugging purposes. 

\section{Experimental Test of the Compact LLRF Prototype for ACCEL}

In the process of prototyping the compact LLRF for ACCEL, a range of experimental tests have been performed for performance evaluation and functional verification purposes. The test begins with the simplest loopback from the DAC to the ADC of the compact LLRF prototype and then test broadens to include the to higher power regime with solid state amplifier and klystron. The test results at different stages of the system or different parts of the system, which are important for the final implementation of the control schemes are summarized and analyzed in this section.

\subsection{\label{loopback} Loopback Test of the Compact LLRF Prototype}

\begin{figure}
\includegraphics[width=3.4in]{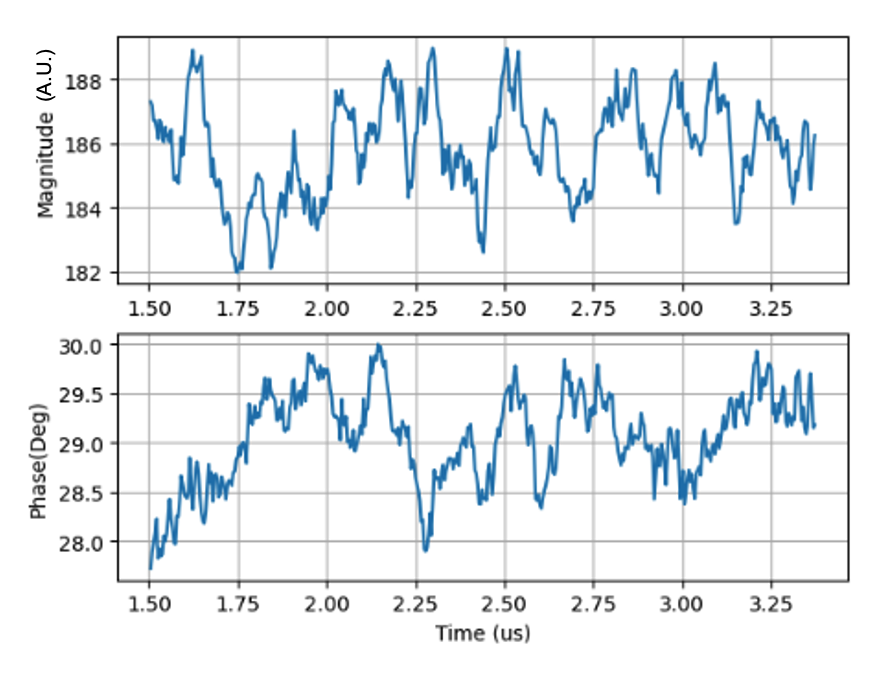} 
\caption{\label{fig:fig-5} The magnitude and phase fluctuation on a 2 \(\mu\)s pulse with DAC amplitude of 8000.}
\end{figure}

\begin{figure}
\includegraphics[width=3.4in]{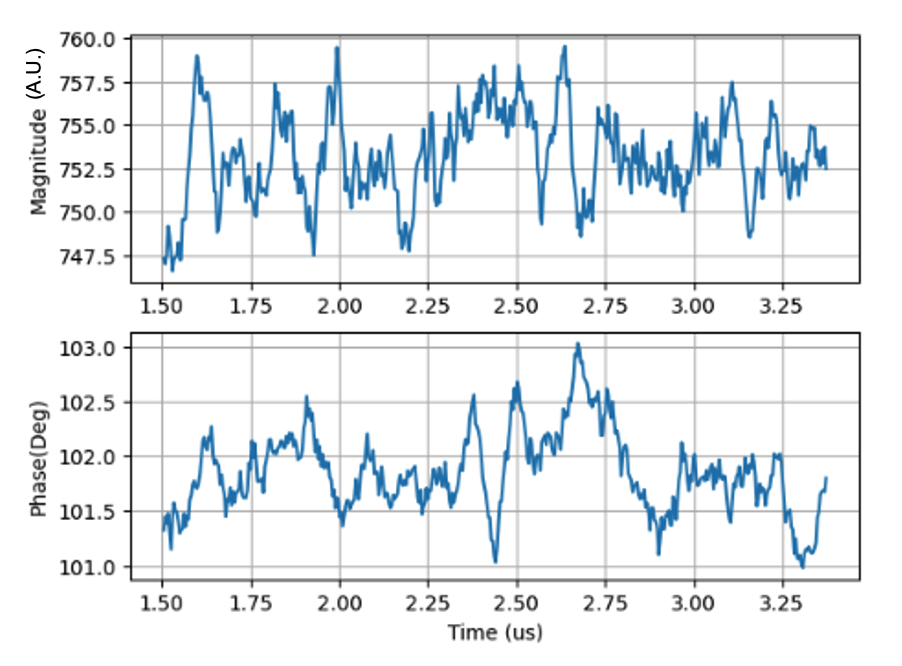} 
\caption{\label{fig:fig-6} The magnitude and phase fluctuation on a 2 \(\mu\)s pulse with DAC amplitude of 32000. \cite{liu2024direct}}
\end{figure}

The direct loopback is a critical test of the compact LLRF system, as it can reveal the performance of the platform without relying on other test equipment. The loopback test can also expose functional defects and performance issues. The direct loopback tests for C-band LLRF system in different test conditions have been summarized in recent publications \cite{liu2024direct, liu2024generationllrfcontrolplatform}. We have performed a compressive direct loopback test, which demonstrated magnitude and phase fluctuation levels of 0.13\% and 0.14\textdegree \, in 1s respectively with both ADC and DAC operating in the optimum input and output ranges. Further testing of magnitude and phase stability is ongoing and will be  available in future publications. In \cite{liu2024direct}, the loopback test with 1 \(\mu\)s pulse at 60 Hz with the a similar SSA and the ZCU216 evaluation board used for tests in this paper has delivered magnitude and phase fluctuation levels of 0.09\% and 0.18\textdegree. The pulse-to-pulse evaluation results are significantly better than the \(\pm\)1\% and \(\pm\)1\textdegree requirements for ACCEL, which means pulse-to-pulse feedback requirement is achievable with the compact LLRF platform. Due to the multi-bunch operation of ACCEL, the flatness of magnitude and phase within each pulse is also required to be \(\pm\)1\% and \(\pm\)1\textdegree. The targeted flat-top duration for each pulse is in 1 to 5 \(\mu\)s range. 

In this case, flatness of the pulse top has been evaluated at two amplitude levels. The RF pulse used for this test is 2 \(\mu\)s at 60 Hz. As the RF pulses are generated by the DAC datapath, the magnitude of base-band pulse is used to denote the level of RF pulse, which named as `DAC amplitude' in this case. The test has been performed at two different DAC amplitude levels, which are 8,000 and 32,000. The 32,000 DAC amplitude is close to the maximum value that can be used for the digital mixing in the pulse generation process. The RF pulse is generated by the DAC of the compact LLRF and then looped back to an ADC via BPFs. The ADC samples the RF pulse directly and IQ components after down conversion are recorded. In the post analysis, the data in IQ format has been converted to magnitude and phase format. The magnitude and phase values on the flat top of the two DAC amplitude levels are shown in Figure \ref{fig:fig-5} and \ref{fig:fig-6}. 

The average values of magnitude measured with DAC amplitude at 8000 and 32000 are approximately 186 and 752, which follows the desired linear trend. There is not obvious common drifts or oscillations for both magnitude and phase values at the two DAC amplitude levels. The magnitude fluctuation on the pulse top is measured by the percentage of standard deviation with respect to average over the flat top. The fluctuation for the 8,000 and 32,000 DAC amplitude levels are 0.8\% and 0.34\%. Phase fluctuation is measured directly with the standard deviation levels, which are 0.47\textdegree \, and  0.37\textdegree \, for the DAC levels of 8,000 and 32,000 respectively. Operating the DAC and ADC in the higher range benefits the fluctuation levels for both magnitude and phase values within a single pulse. The amplification or attenuation levels in the final LLRF system should be optimized with respect to the input and output ranges of the ADCs and DACs. Although the RF signal used in this test is on the lower side of the ADC, the flatness levels achieved are still considerably better than the requirement of 1\% and 1\textdegree \, which indicates that the compact LLRF prototype can generate and measure the RF pulses within the desired fluctuation level.We have analyzed the fluctuations on the pulse top in frequency domain, but no consistent periodic disturbance was identified. 

\subsection{\label{tuning} RF Frequency Tuning Scheme Tests}

As introduced in Section \ref{sec:rf_freq}, the ACCEL is required to operate in a wide operating temperature range, and there is no temperature stabilization for the accelerating structure. The resonant frequency of the structure changes with temperature. Therefore, an RF frequency tuning scheme has been designed to track the resonant frequency of the accelerating structure and align the operating RF frequency with the resonance.  In this section, the technique for fine frequency tuning will be presented and the tuning routine implemented will be tested with both single-cell and dual-cell ACCEL accelerating structures.

\subsubsection{\label{tuning_single}Single-cavity Structure Test}

\begin{figure}
\includegraphics[width=3.4in]{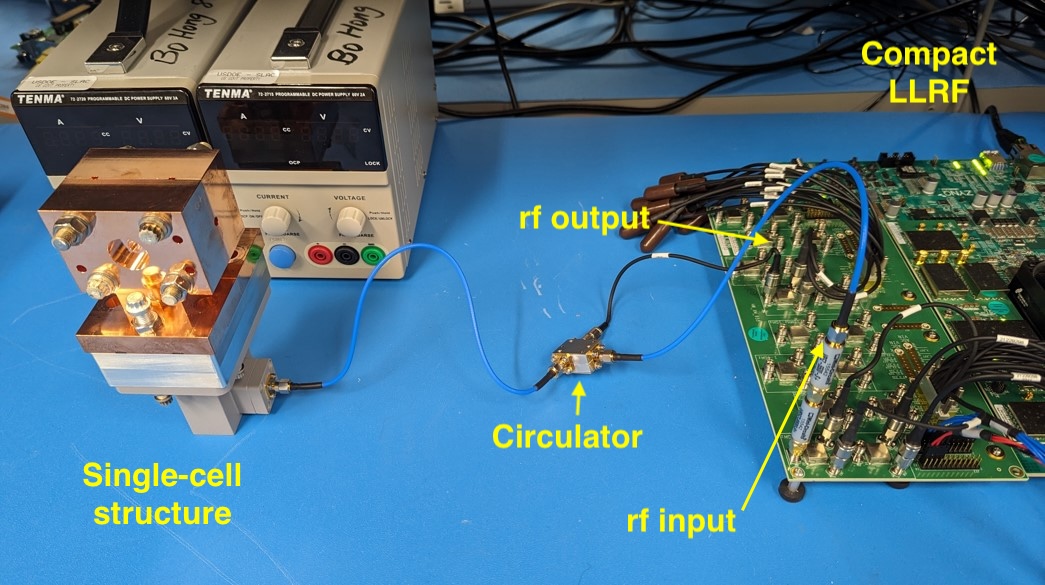} 
\caption{\label{fig:fig-7} The test setup of single-cell structure with the compact LLRF for ACCEL.}
\end{figure}

The first test of RF frequency tuning was performed with the simple test setup shown in Figure \ref{fig:fig-7}. The RF output of the compact LLRF system generates RF pulses at 60 Hz. The RF pulses are injected into the single-cell structure, which is a prototype cavity for ACCEL, via a circulator. The circulator with a bandwidth of 4 to 8 GHz, PACL1704000800A, manufactured by DBwave Technologies LLC has been used for this test. The circulator couples the reflected RF signal from the structure, and the signal is connected to one of the RF input channels of the compact LLRF system. The cavity reflection signal is sampled directly by the ADC channel and down converted by the following datapath. The IQ samples collected are converted to magnitude and phase for analysis in this section.

\begin{figure}
\includegraphics[width=3.4in]{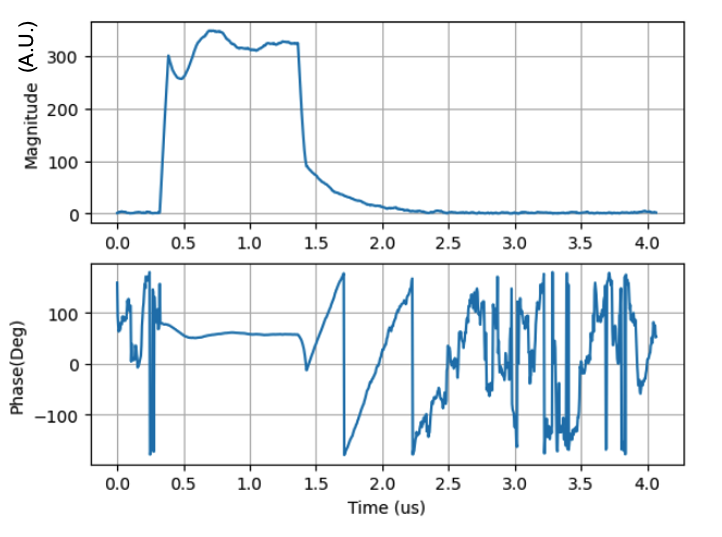} 
\caption{\label{fig:fig-8} The magnitude and phase values of cavity reflection signal before the RF frequency tuning.}
\end{figure}

For the first step, the test setup was driven by RF pulses with the RF frequency of 5.712 GHz and measured at the same RF frequency. The magnitude and phase values of the cavity reflection signal are shown in Figure \ref{fig:fig-8}. The magnitude of the cavity signal has ripples after the initial rise and remains generally flat until the RF pulse is off. The reflection signal remains high over the pulse duration, indicating that the RF power injected to the structure is almost fully reflected as a result of driving the system off the resonant frequency of the cavity. When the structure is driven off resonance within certain threshold, the structure will still be energized, but with an extremely low gain. The single-cell structure has been designed to resonant at 5.712 GHz, but machining errors and ambient temperature can introduce offsets to its resonance. The phase level of the reflection signal remains stable until the RF pulse is off and there is a steep linear ramp after the RF is off. The reflection after the RF pulse is off is dominated by the radiation from the structure cavity. The structure radiates the RF energy at its resonance frequency. When the RF frequency and the down conversion frequency are different, it will be reflected as a linear ramp on the phase level. The offset between the structure resonance and the operation RF frequency of the LLRF system can be calculated with Equation \ref{equ:1}. 

\begin{equation}
 \Delta\omega = \frac{d\phi}{dt}
\label{equ:1}
\end{equation}

\begin{figure}
\includegraphics[width=3.4in]{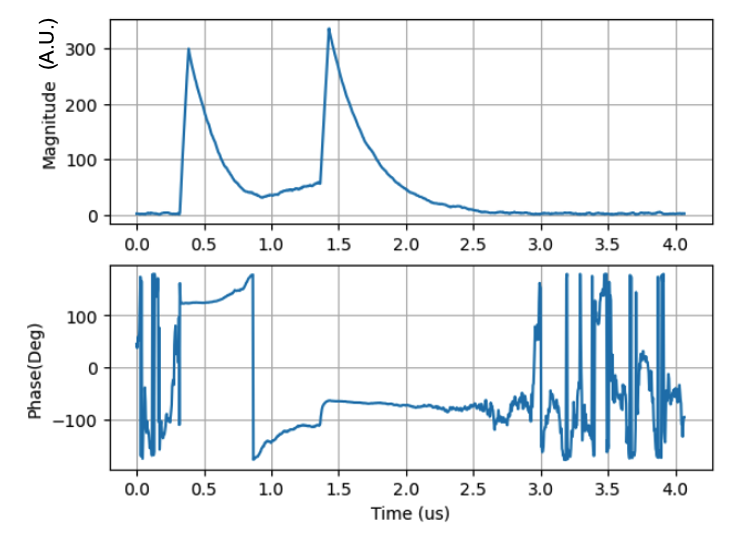} 
\caption{\label{fig:fig-9} The magnitude and phase values of cavity reflection signal after the RF frequency tuning.}
\end{figure}

The phase values has been transferred to software and frequency offset has been computed with the slope of phase ramp after the RF is off. The resonant frequency of the single-cell structure computed using Equation \ref{equ:1} is approximately 2.08 MHz higher than 5.712 GHz. Then the RF frequency of the LLRF has been adjusted to the resonance frequency of the cavity, and the procedure above has been repeated. The magnitude and phase levels of the cavity reflection signal after frequency tuning are shown in Figure \ref{fig:fig-9}. The magnitude of the reflection ramps down after the first spike and that demonstrates the field filling process in the accelerating structure. The cavity reflection reaches the lowest magnitude when the filling process is complete. As there is no beam and the structure is overcoupled, the second peak on the magnitude of reflection is higher than the first one. The filling and dissipation processes also reflect on the phase level. The phase has been reversed when the structure is fully filled. The magnitude and phase levels before the RF pulse is switched off show that the RF signal can be successfully injected to the structure and established field to fill the structure. 

After the RF pulse is off, the magnitude ramps down, which shows the stored RF power radiating and dissipating in the structure. The phase level remains almost flat after the pulse is off, which means that the operation frequency of the LLRF is close to the resonance frequency of the structure. That signifies that the RF frequency tuning has been successful for the single-cell structure with high precision.

\subsubsection{Dual-cavity Structure Test}

\begin{figure*}
\includegraphics[width=7 in]{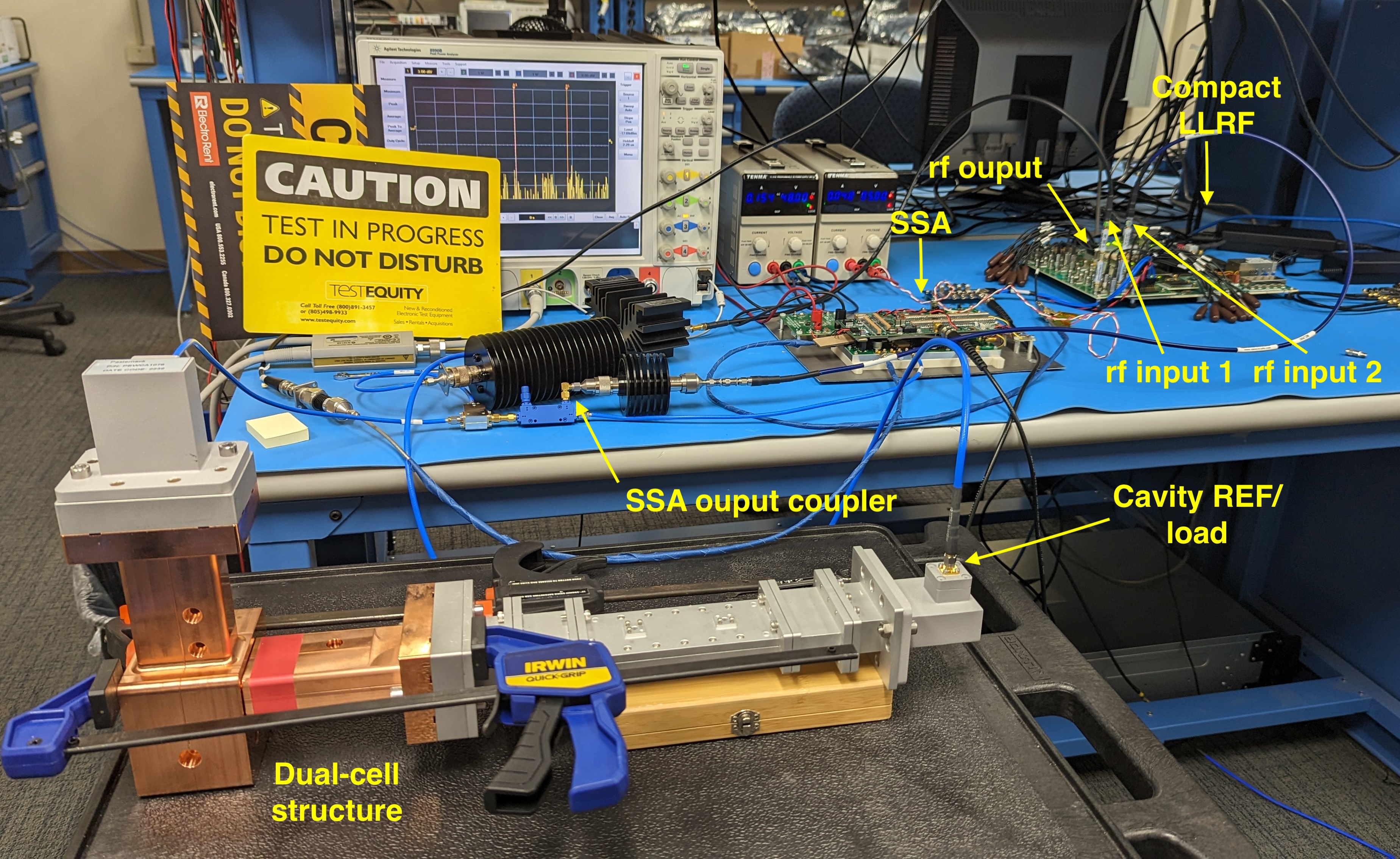}%
\caption{\label{fig:fig-10} The test setup of the two-cell structure with the SSA and the compact LLRF prototype. }
\end{figure*}

In the final design of ACCEL, the full accelerating structure will be composed by 26 dual-cell structures. After the RF frequency tuning scheme has been verified to be functional with the single-cell structure, the test has been extended to the dual-cell structure with a higher power SSA. Figure \ref{fig:fig-10} shows the test setup with the dual-cavity structure. The RF pulse generated by the compact LLRF system is amplified by a low noise amplifier, as the output power of the compact LLRF is too low to drive the SSA. The SSA amplifies the resultant RF pulse again, and the peak power is up to 300 W. The forward RF power is measured at the output of the SSA via a coupler. The output of the SSA is injected into a dual-cell structure via an adapter. The reflection signal is measured at the reflection port through a load and an adapter. Both the forward and the reflection signals are attenuated and looped back via the RF input measurement channels.

\begin{figure}
\includegraphics[width=3.4in]{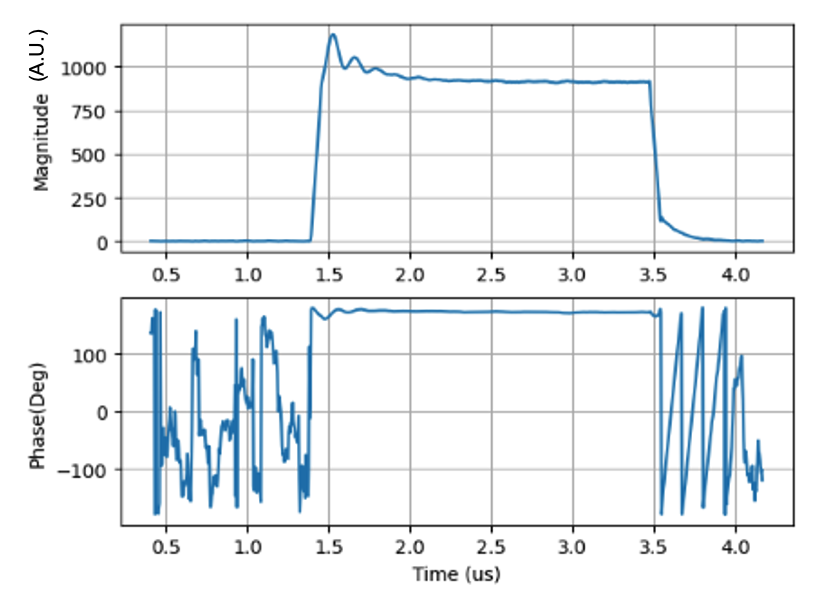} 
\caption{\label{fig:fig-11} The magnitude and phase values of the cavity reflection signal before the RF frequency tuning.}
\end{figure}

\begin{figure}
\includegraphics[width=3.4in]{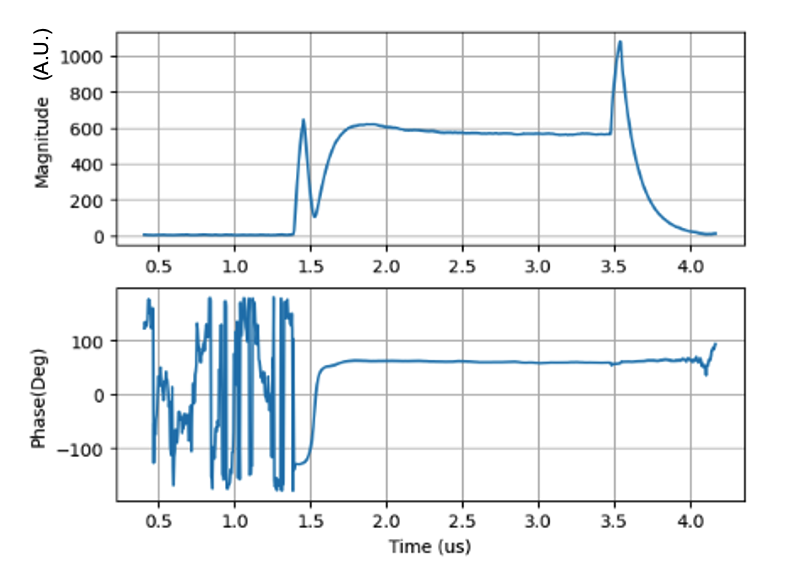}
\caption{\label{fig:fig-12} The magnitude and phase values of cavity reflection signal after the RF frequency tuning.}
\end{figure}

Compared with the single-cell structure, the RF frequency tuning of the dual-cell structure is more complicated. As the two cells of the structure have different resonance frequencies due to the machining tolerance. The resonant frequencies of the cells are measured using the same method described in Section \ref{tuning_single} with one of the cells detuned. The measured resonant frequencies of the two cells are differed by approximately 0.32 MHz. The resonance frequency for the one of the cell is then tuned close to the other cell by inserting a metal wire via the hole on the wall of structure. For the dual-cell structure in production, this step will be replaced by mechanical tuning before assembly. 

After the resonance frequency of the two cells are tuned to a similar level, the RF frequency tuning scheme was tested. The test begins with RF pulse at RF frequency of 5.712 GHz and Figure \ref{fig:fig-11} shows the magnitude and phase levels of the cavity reflection signal. The magnitude of the reflection oscillates with a slightly reducing magnitude after the initial rise, which indicates that the power is almost fully reflected and the structure has been energies with a extremely low gain. Then the magnitude level settle in about 0.6 \(\mu\)s showing that no more power has been injected to the structure afterwards. While the pulse is on, the phase of the reflection remains in a stable level with some oscillation at early stage. After RF pulse is switched off, there is linear ramp on the phase level with steep slop. The resonant frequency computed for the for the dual-cell structure is around 7.5 MHz higher than 5.712 GHz.

The test was performed again after the operation RF frequency of the compact LLRF system is tuned to the structure resonant frequency and the magnitude and phase levels of the cavity reflection signal are shown in Figure \ref{fig:fig-12}. The magnitude of the reflection signal shows a different field filling process compared with the single-cell structure. The dual-cell structure has been designed to have a coupling factor (\(\beta\)) around 4.6 and the magnitude is expected to reach the minimum after the first peak in 100 ns. The magnitude of the cavity reflection shown in the figure reaches the minimum value around 100 ns as the radiation from the cavity and reflection from the iris almost canceled with each other. Then the structure cells are filled for about 500 ns until the reflection signal reaches an steady level. As the structure is designed to be overcoupled without the beam, the second peak after the RF is switched off is approximately 60 \% higher than the first peak. The phase remains flat after the RF pulse is off, indicating that the operating frequency of the prototype ACCEL compact LLRF is extremely close to the resonance of the dual-cell structure. This shows that the RF frequency tuning scheme of the LLRF system can successfully measure the resonant frequency of the dual-cell accelerating structure and the tune operating frequency to be aligned with the cavity resonance frequency.  

\subsection{\label{highpower} Compact LLRF Prototype Tests with High-power Test Facility}

\begin{figure}
\includegraphics[width=3.4in]{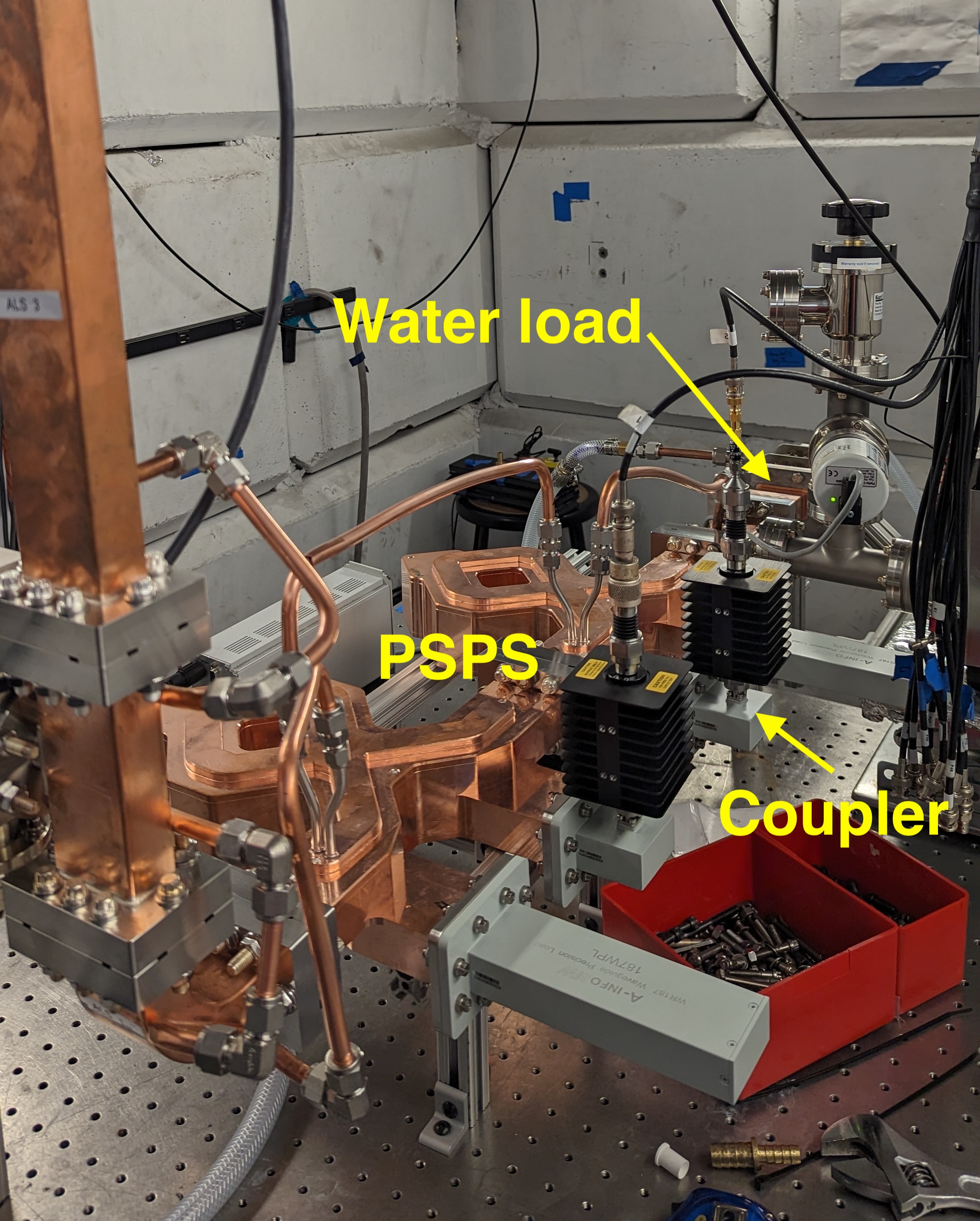} 
\caption{\label{fig:fig-13} The test setup of single-cell structure with the compact LLRF for ACCEL.}
\end{figure}

\begin{figure}
\includegraphics[width=3.4in]{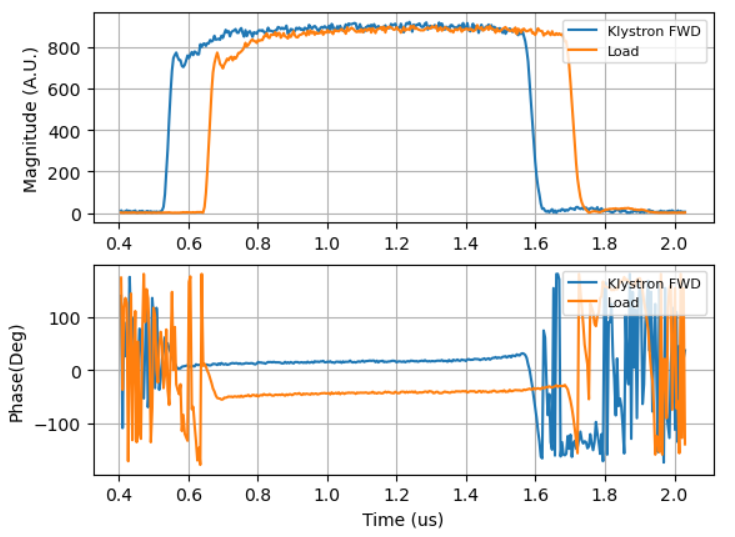} 
\caption{\label{fig:fig-14} The magnitude and phase of the klystron forward and load forward signal measured in high-power test.}
\end{figure}

\begin{figure}
\includegraphics[width=3.4in]{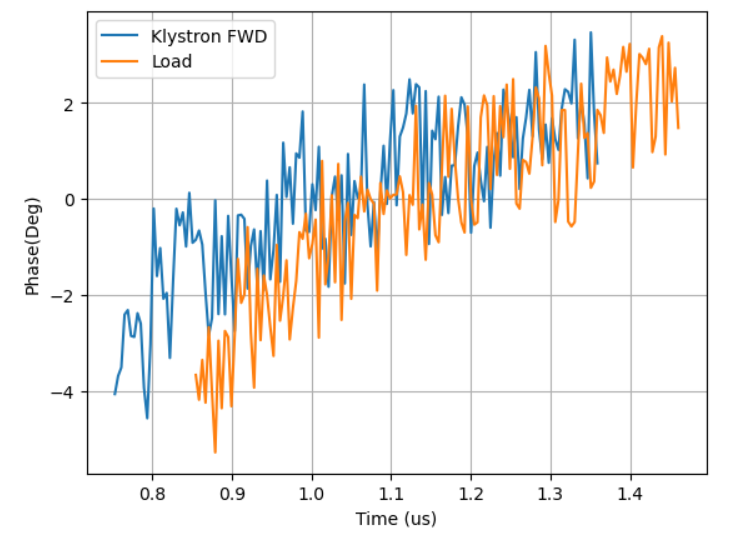} 
\caption{\label{fig:fig-15}  The phase levels of the klystron forward and load forward signals on the 600 ns flat top measured in high-power test.}
\end{figure}

The compact LLRF platform was tested at the C-band test facility at Radiabeam at higher power levels. The test was focused on investigating the fluctuation levels and patterns on the klystron forward signal, which is critical for the design and implementation of pulse shape control scheme. The C-band klystron is driven by the RF pulse generated by the RF output of the compact LLRF platform. Then the output RF power of klystron is coupled to the test bunker via a waveguide. For this test, the peak power generated by the klystron is 426 kW and 199 kW is delivered to the load in the test bunker. Figure \ref{fig:fig-13} shows the setup in the test bunker. The RF power from the waveguide is coupled to a water load via a power splitter and phase shifter (PSPS). 

The RF power at the output of the klystron and the input of water load are measured with directional couplers. Figure \ref{fig:fig-14} shows the magnitude and phase levels of the klystron forward and load forward, which are labeled as 'klystron FWD' and 'Load' respectively. The magnitude of the first peak of the load forward signal is scaled to match the first peak of the klystron forward signal. The magnitude and phase levels of the load follow the klystron forward power as expected and both have a ramp up about 100 ns after the first peak. The flat top with steady magnitude and phase is approximately 600 ns within the 1 \(\mu\)s RF pulse.

The magnitude fluctuation is evaluated with the percentage of the standard deviation of the magnitude values in the 600 ns flat top with respect to the average magnitude in the period. The magnitude fluctuation levels for the klystron forward and load forward are  1.60 \% and 1.51 \% respectively. The phase fluctuation level is measured directly with the standard deviation of the phase values on the flat top. The phase fluctuation levels for the klystron forward and load forward are  1.66\textdegree \, and 1.98\textdegree \, respectively. The magnitude and phase fluctuation levels at klystron forward and load forward are consistent, which means the RF power generated by the klystron has been successfully delivered to the load in the test bunker and RF signals are measured with high precision by the compact LLRF platform. The fluctuation levels for both magnitude and phase measured on the flat top are above the 1\% and 1\textdegree \, requirement of ACCEL, which verified the necessity of the pulse shape control scheme. 

Figure \ref{fig:fig-15} shows the phase levels of the klystron forward and load forward signals on the 600 ns flat top. The average phase of each of the signals has been subtracted. There is a linear phase ramp starting from approximately -4\textdegree \, and up to 3\textdegree \, until the end of the flat top on both of the signal. The ACCEL system requires a flat top between 1 to 5 \(\mu\)s, so the phase drift will be even more significant. The phase drift introduced by the klystron is much higher the ACCEL requirement. The drift needs to be compensated for by the pulse shape control scheme with phase modulation techniques. As the modulation of the compact LLRF is fully implemented in the digital domain, the phase and magnitude modulation can be implemented with high flexibility. We have tested a similar RFSoC based LLRF with several phase modulation schemes at high power regime. The high-power test demonstrated the capability to generate and measure phase modulation schemes with high precision, and the full results will be published in the near future. 

\section{Conclusion}

A compact LLRF system prototype based on RFSoC technology has been designed and developed for the ACCEL program. Compared with conventional LLRF systems, the higher integration level and direct RF sampling of RFSoC offer distinctive SWaP advantages for the compact LLRF system we proposed for ACCEL. Due to the special operation and application requirements of ACCEL, additional control schemes, such as automated RF frequency tuning and pulse shape control, are required. The initial control schemes proposed, including: the time sequence, the building blocks, and the interface between software and firmware have been introduced. The hardware, firmware and software of the compact LLRF platform have been designed, implemented and tested, with each block at different levels of completion. The development of the complete prototype will continue in the next phase of the project.

The prototype enabled a range of feasibility studies and experimental tests for different purposes. In the initial loopback test, the compact LLRF system demonstrated magnitude and phase fluctuation levels at 0.34\% and 0.37\textdegree \, on the flat top of a 2 \(\mu\)s RF pulse. The RF frequency tuning scheme test demonstrates that the scheme can measure the resonant frequency of single-cell and dual-cell ACCEL accelerating structure prototypes and tune the operation RF frequency of the LLRF system with high precision. The field filling, reflecting, and dissipation processes can be visualized on the reflection signal from the structure before and after the RF frequency is tuned. 

The high-power test has been performed in the lower power regime of the klystron, which has a maximum peak power output of 50 MW. As the klystron was operated far below saturation,  the RF power generated is less stable. The klystron forward and load forward measurements revealed fluctuation levels and drift patterns introduced by the the klystron. Those features will be a critical reference when implementing the full pulse shape control scheme. 

There are a range of test and development tasks that must be completed before the compact LLRF can be operated with all the required control schemes. The firmware building blocks of the pulse-to-pulse feedback control scheme have been implemented, but not yet tested in a full-firmware implementation. In this phase of the project, we did not have enough resources to perform high power tests with the ACCEL prototype and more development effort is required to demonstrate feedback correction. For the next phase of the project, we plan to start the development by demonstrating a simple loopback that drives the structure, measures the RF field, and adjusts the pulse with respect to a user defined set-point. Once we have some test data, the feedback loop design can be tuned based on it. Then the complete compact LLRF platform can be tested with the ACCEL prototype structure in higher power regime.

\section*{Acknowledgment}

 The work of the authors is supported by the DARPA. The views, opinions and/or findings expressed are those of the authors and should not be interpreted as representing the official views or policies of the Department of Defense or the U.S. Government. The work of the authors is also partially supported by the U.S. Department of Energy under Contract No. DE-AC02-76SF00515.

\section*{Data Availability Statement}

The data underlying this article will be shared on reasonable request to the corresponding author.

\nocite{*}
\bibliography{bibliography}

\end{document}